\documentclass[12pt,a4paper]{article}
\usepackage{citesort}
\usepackage{a4wide}
\usepackage{amsmath}
\usepackage{amssymb}
\usepackage{epsfig}
\usepackage{subfigure}
\usepackage{exscale}
\usepackage{float}
\usepackage{bbm}
\usepackage[numbers,sort&compress]{natbib}
\usepackage{pst-plot, pstricks,pst-math}
\usepackage{fancybox,amssymb,color}
\usepackage{graphicx}
\usepackage{pstricks, color, graphicx, epsfig, psfrag}
\usepackage{amsfonts,amsmath,amssymb,slashed}
\usepackage{dsfont}
\usepackage{bbm,bm}
\usepackage{fancyhdr, a4wide}
\usepackage[english]{babel}
\usepackage{subfigure}

\newcommand{\Dag}{^{\dagger}}
\newcommand{\up}{\uparrow}
\newcommand{\down}{\downarrow}
\newcommand{\Tr}{\mbox{Tr}}
\newcommand{\Lag}{\mathcal{L}}

\makeatletter
\@addtoreset{equation}{section}
\makeatother

\title{Microscopic Model versus Systematic Low-Energy Effective Field Theory
for a Doped Quantum Ferromagnet}

\author{U.\ Gerber$^a$, C.\ P.\ Hofmann$^b$, F.~K\"ampfer$^c$, and 
U.-J.~Wiese$^a$
\\ \\
\normalsize {$^a$ Center for Research and Education in Fundamental Physics} \\
\normalsize {Institute for Theoretical Physics, Bern University} \\
\vspace{0.3cm}
\normalsize {Sidlerstrasse 5, CH-3012 Bern, Switzerland} \\
\normalsize {$^b$ Facultad de Ciencias, Universidad de Colima} \\
\vspace{0.3cm}
\normalsize {Bernal D\'iaz del Castillo 340, Colima C.P.\ 28045, Mexico} \\
\normalsize {$^c$ Condensed Matter Theory Group, Department of Physics} \\
\normalsize {Massachusetts Institute of Technology (MIT)} \\
\normalsize {77 Massachusetts Avenue, Cambridge, MA 02139, U.S.A.} \\}

\vspace{-1truecm}

\begin{document} 
\maketitle

\vspace{-1truecm}

\begin{abstract} \normalsize

We consider a microscopic model for a doped quantum ferromagnet as a test case 
for the systematic low-energy effective field theory for magnons and holes, 
which is constructed in complete analogy to the case of quantum 
antiferromagnets. In contrast to antiferromagnets, for which the effective 
field theory approach can be tested only numerically, in the ferromagnetic case
both the microscopic and the effective theory can be solved analytically.
In this way the low-energy parameters of the effective theory are determined
exactly by matching to the underlying microscopic model. The low-energy
behavior at half-filling as well as in the single- and two-hole sectors is
described exactly by the systematic low-energy effective field theory. In
particular, for weakly bound two-hole states the effective field theory even
works beyond perturbation theory. This lends strong support to the quantitative
success of the systematic low-energy effective field theory method not only in
the ferromagnetic but also in the physically most interesting antiferromagnetic
case.

\end{abstract}
 
\maketitle
 
\newpage

\section{Introduction}

Achieving a quantitative understanding of the doped antiferromagnetic
precursors of high-temperature superconductors is a great challenge in
condensed matter physics. In particular, away from half-filling Monte Carlo
simulations of these strongly correlated electron systems suffer from a very
severe sign problem. Also analytic calculations in underlying microscopic
Hubbard or $t$-$J$-type models are not fully systematic but suffer
from uncontrolled approximations. Particle physicists face similar challenges
in the physics of the strong interactions between quarks and gluons.
Remarkably, the low-energy physics of pions --- the pseudo-Goldstone bosons of
the spontaneously broken $SU(2)_L \times SU(2)_R$ chiral symmetry of QCD ---
is described quantitatively by a systematic effective field theory
\cite{Col69,Cal69,Wei79,Gas85}, known as chiral perturbation theory.
Similarly, the low-energy physics of the spin waves or magnons --- the
Goldstone bosons of the spontaneously broken $SU(2)_s$ spin symmetry in an
antiferromagnet --- is also captured by a systematic effective field theory
\cite{Cha89,Neu89,Fis89,Has90,Has91,Has93,Chu94}. Early attempts to include
doped holes into the effective description of antiferromagnets are described
in \cite{Shr88,Wen89,Sha90,Sus04}. Motivated by the quantitative success of
baryon chiral perturbation theory \cite{Gas88,Jen91,Ber92,Bec99} for pions and 
nucleons in QCD, fully systematic low-energy effective field theories have been
developed for hole-doped antiferromagnets both on a square \cite{Kae05,Bru06} 
and on a honeycomb lattice \cite{Jia09}, as well as for electron-doped 
antiferromagnets on a square lattice \cite{Bru07}. The resulting systematic 
effective field theories have been used to study magnon-mediated two-hole 
\cite{Bru06,Bru06a} and two-electron bound states \cite{Bru07} as well as 
spiral phases in the staggered magnetization order parameter 
\cite{Bru07,Bru07a,Jia09}. The quantitative correctness of the magnon effective
field theory has been demonstrated in great detail at permille level accuracy 
by comparison with Monte Carlo simulations of the quantum Heisenberg model 
using the very efficient loop-cluster algorithm \cite{Wie94,Bea96,Ger09}.
Similarly, the single-hole sector of the $t$-$J$ model has been simulated both
on the square \cite{Bru00,Mis01} and on the honeycomb lattice \cite{Jia08a}. 
Indeed, the observed location of the hole pockets in the Brillouin zone has 
provided important input for the construction of the various systematic 
effective field theories for doped antiferromagnets.

In general, low-energy effective field theories cannot be derived rigorously
from the underlying microscopic physics. Instead one performs a detailed 
symmetry analysis of the underlying theory and constructs all terms in the
effective Lagrangian that are invariant, order by order in a systematic
derivative expansion. Each term is then endowed with an a priori undetermined
low-energy parameter. In particular, the values of these parameters are not
fixed by symmetry considerations, but must be determined by matching to the
underlying microscopic system. This can be done by comparison with either
experiment or numerical simulations. Only in exceptional cases the
underlying microscopic model can be solved analytically and the low-energy
parameters can be determined exactly. One such case is the ferromagnetic
Heisenberg model whose low-energy physics was analytically derived by Dyson
\cite{Dys56}. The corresponding low-energy effective theory was constructed by
Leutwyler \cite{Leu94} and discussed in great detail in
\cite{Hof99,Rom99,Bae04}. Remarkably, in contrast to the effective theory for
antiferromagnets, the effective theory for ferromagnets contains an additional
Wess-Zumino term whose quantized prefactor is the total magnetization. The
values of the magnetization and of the spin stiffness --- the other leading
order low-energy parameter of a ferromagnet --- can be easily read off from
Dyson's analytic solution of the underlying microscopic Heisenberg model.
Thanks to the analytic solvability of the ferromagnetic Heisenberg model, in
this case the predictions of the effective theory can be verified rigorously.
Indeed, once the low-energy parameters have been fixed by matching to the
underlying system, in the low-energy domain the effective theory yields
exactly the same results as the Heisenberg model. It is interesting to note
that the calculations in the effective theory are much simpler than those in
the microscopic model. 

The effective field theories for doped antiferromagnets mentioned above have
again been constructed based on symmetry considerations. However, in that
case the underlying 2-dimensional Hubbard or $t$-$J$-type models cannot be 
solved analytically, and one must hence rely on numerical methods for fixing
the low-energy parameters and for verifying the validity of the low-energy
effective theory. In this paper, we consider a microscopic Hubbard-type model
for a doped ferromagnet which can be solved analytically. Furthermore, the
corresponding low-energy effective theory can be constructed in exactly the
same way as in the antiferromagnetic case. By showing explicitly that the
microscopic and the effective theory of the doped ferromagnet yield identical
results, we lend  further support to the general construction principle for
the effective theories. For simplicity, our analytic study will be performed
in one spatial dimension, but the extension to higher dimensions is
straightforward. It should be noted that in one spatial dimension the
antiferromagnetic Heisenberg model is analytically solvable by the Bethe
ansatz \cite{Bet31}. According to Haldane's conjecture \cite{Hal83}, the
corresponding low-energy effective theory is a 2-dimensional $O(3)$ non-linear
$\sigma$-model at vacuum angle $\theta = \pi$. As first noted by Lieb and Wu,
in one dimension even the Hubbard model can be solved analytically
\cite{Lie68,Lie03}. In particular, these authors have shown that this model
has no Mott transition. We prefer to consider the ferromagnetic model because
it is easier to solve analytically and because its low-energy effective theory
is similar to the one of the doped antiferromagnets. It should be pointed out
that our ferromagnetic model is not meant to provide a realistic description
of ferromagnetism in actual materials. This would require two bands as well as
Hund rule couplings \cite{Fro05}. Instead, for simplicity, we impose
ferromagnetism by including the corresponding coupling by hand. Still, the
range of applicability of the effective theory to be constructed in this paper
goes beyond our ferromagnetic model, as the effective theory applies to any
system exhibiting the same symmetries and symmetry breaking pattern as the
microscopic model considered here. 

Systematic low-energy effective field theories have also been used in studying
light nuclei
\cite{Wei90,Kap98,Epe98,Bed98,Kol99,Epe01,Bea02,Bed02,Nog05,Bea08}.
In this case, due to nuclear binding, non-perturbative effects must be
understood in the framework of the low-energy effective theory. Currently,
there is still discussion about how this can be achieved completely
systematically. Just as light nuclei are bound states of a few nucleons, the
doped ferromagnet studied in this paper develops bound states of holes.
Interestingly, their dynamics can be understood analytically both in the
underlying and in the effective theory. Hence, the doped ferromagnet is a
system in which systematic approaches to non-perturbative problems in
effective field theory can be tested. Thus, the investigations in this or
related models may also have an impact on the corresponding issues arising in
the context of the strong interactions.

The paper is organized as follows. In section 2 the underlying microscopic
model is introduced and its symmetry properties are investigated in detail.
The model is then solved at half-filling, as well as in the one- and two-hole
sectors. In particular, the dispersion relations of magnons and holes, as well
as the binding energy of two holes and the two-hole scattering states are
determined analytically. In section 3 the corresponding low-energy effective
field theory is constructed using the non-linear realization of the
spontaneously broken $SU(2)_s$ spin symmetry. In particular, the hole fields are
included in the same way as for a doped antiferromagnet. In section 4 magnons,
single holes, as well as two-hole scattering and two-hole bound states are
investigated in the effective field theory framework. The a priori undetermined
low-energy parameters are fixed by matching to the underlying microscopic
system, and it is verified explicitly that the predictions of the effective 
theory agree exactly with those of the microscopic model. Finally, section 5 
contains our conclusions. Some technical details are presented in an appendix.

\section{Construction and Solution of a Microscopic Model for a
Doped Ferromagnet}

In this section we construct a Hubbard-type microscopic model for a doped
ferromagnet, investigate its symmetries, and then solve it in the zero-, one-,
and two-hole sectors.

\subsection{Microscopic Model for Ferromagnetism}

Let us construct a microscopic model describing the hopping of fermions on a
1-dimensional lattice with spacing $a$, with the Hamiltonian
\begin{equation} 
\label{hamil}
H = - t \sum_x \left(c_x\Dag c_{x+a} + c_{x+a}\Dag c_x\right) -
J \sum_x \vec S_x \cdot \vec S_{x+a} + 
\frac{U}{2} \sum_x (c_x\Dag c_x - 1)^2.
\end{equation}
The creation and annihilation operators for fermions at a site $x = a n, \
n \in \mathbb{Z}$, with spin $s = \up, \down$ are given by
\begin{equation}
c_x\Dag = \left(c_{x\up}\Dag, c_{x\down}\Dag\right), \quad 
c_x = \left(\begin{array}{c} c_{x\up} \\ c_{x\down} \end{array} \right).
\end{equation}
They obey the standard anticommutation relations
\begin{equation}
\left\lbrace c_{x s}\Dag,c_{x' s'} \right\rbrace = \delta_{xx'}\delta_{ss'}, 
\quad 
\left\lbrace c_{x s},c_{x' s'} \right\rbrace = 
\left\lbrace c_{x s}\Dag,c_{x' s'}\Dag \right\rbrace = 0.
\end{equation}
Putting $\hbar = 1$, the spin operator at the site $x$ is given by
\begin{equation}
\vec S_x = c_x\Dag \frac{\vec \sigma}{2} c_x,
\end{equation}
where $\vec \sigma$ denotes the Pauli matrices. Let us discuss the
various terms in
the Hamiltonian above. The term proportional to $t$ describes hopping of
fermions by one lattice spacing, i.e.\ it represents the kinetic energy. The
parameter $J>0$ is a ferromagnetic exchange coupling constant, while the term
proportional to $U>0$ describes an on-site Coulomb repulsion. As mentioned
earlier, this Hamiltonian does not provide a realistic description of real
ferromagnetic materials. We consider it because it is analytically solvable at
low energies and can thus be used to test the corresponding effective theory.

The microscopic model defined by eq.(\ref{hamil}) has various symmetries, which
we are going to discuss now. It is straightforward to confirm that the 
Hamiltonian commutes with the total spin
\begin{equation}
[H,\vec S] = 0, \quad \vec S = \sum_x \vec S_x,
\end{equation}
and is thus invariant under global $SU(2)_s$ spin rotations. As we will see 
later, the $SU(2)_s$ symmetry is spontaneously broken down to the subgroup
$U(1)_s$ by the formation of a uniform magnetization. It should be noted that
this is not in contradiction with the Mermin-Wagner theorem. The generators of
another symmetry --- a non-Abelian $SU(2)_Q$ extension of the Abelian $U(1)_Q$
fermion number \cite{Zha90,Yan90} --- are given by
\begin{equation} \label{qgenerators}
Q^+ = \sum_x (-1)^{x/a} c_{x\up}\Dag c_{x\down}\Dag, \quad 
Q^- = \sum_x (-1)^{x/a} c_{x\down} c_{x\up}, \quad 
Q^3 = \sum_x \frac{1}{2} (c_x\Dag c_x - 1).
\end{equation}
The factor $(-1)^{x/a}$ distinguishes between the two sublattices $A$ and $B$ of
even and odd sites. Unlike for an antiferromagnet, it may seem unnatural to
make such a distinction for a ferromagnet. However, as we will see later on,
the introduction of two sublattices is also important for a ferromagnet, as it
will allow us to correctly identify the transformation properties holes and
electrons in the effective theory. It is straightforward to convince oneself
that the Hamiltonian is indeed invariant, i.e.\ 
\begin{equation}
[H,\vec Q] = 0, \quad \vec Q =(Q^1,Q^2,Q^3), \quad Q^\pm = Q^1 \pm i Q^2.
\end{equation} 
It should be pointed out that the $SU(2)_Q$ symmetry would be explicitly broken 
down to $U(1)_Q$ if hopping terms between sites belonging to the same 
sublattice would be included in the Hamiltonian. Furthermore, it is worth
noting that the generators of $SU(2)_s$ commute with those of $SU(2)_Q$.

A displacement $D$ by one lattice spacing is generated by the unitary operator 
$D$ which acts as
\begin{equation}
^Dc_x = D\Dag c_x D = c_{x + a}.
\end{equation}
By relabeling the sum over the lattice points, it is easy to show that
$[H,D] = 0$.
Another discrete symmetry is the spatial reflection $R$, which acts as
\begin{equation}
^Rc_x = R\Dag c_x R = c_{-x}.
\end{equation}
Again, by relabeling the sum over the lattice points, it follows that
$[H,R] = 0$.
Another important symmetry is time reversal which is implemented by an 
anti-unitary operator $T$.

It is useful to introduce a matrix-valued fermion operator
\begin{equation}
C_x^A = \begin{pmatrix} c_{x\up} & c_{x\down}\Dag \\ c_{x\down} & - c_{x\up}\Dag 
\end{pmatrix}, \quad x \in A, \quad
C_x^B = \begin{pmatrix} c_{x\up} & - c_{x\down}\Dag \\ c_{x\down} & c_{x\up}\Dag 
\end{pmatrix}, \quad  x \in B.
\end{equation}
Under combined transformations $g \in SU(2)_s$ and $\Omega \in SU(2)_Q$ it 
transforms as
\begin{equation} \label{F_transform}
^{\vec Q} C_x' = g C_x \Omega^T.
\end{equation}
Under the displacement symmetry one obtains
\begin{equation}
^D C_x^A = C_{x + a}^B \sigma_3, \quad ^D C_x^B = C_{x + a}^A \sigma_3.
\end{equation}
The appearance of the Pauli matrix $\sigma_3$  is due to the factor 
$(-1)^{x/a}$. Under the spatial reflection $R$, which turns $x$ into $Rx=-x$, 
one obtains
\begin{equation}
^R C_x = C_{-x}.
\end{equation}
The Hamiltonian can now be expressed in a manifestly $SU(2)_s$-, $SU(2)_Q$-, 
$D$-, and $R$-invariant form
\begin{eqnarray} 
\label{H_nice}
H&=&- \frac{t}{2} \sum_x \Tr \left[C_x\Dag C_{x + a} + C_{x + a}\Dag C_x\right] -
\frac{J}{16} \sum_x \Tr \left[C_x\Dag \vec\sigma C_x\right] \cdot 
\Tr \left[C_{x + a}\Dag \vec\sigma C_{x + a}\right] \nonumber \\ 
&+&\frac{U}{12} \sum_x \Tr \left[C_x\Dag C_x C_x\Dag C_x \right].
\end{eqnarray}

\subsection{Eigenstates for Electrons and Holes}

We will now construct electron and hole states above a half-filled ground state
containing up-spin fermions at each lattice site. The corresponding vacuum
state is given by 
\begin{equation}
|v\rangle = \prod_x c_{x\up}\Dag |0\rangle,
\end{equation}
where $|0\rangle$ represents an empty lattice without any fermions. Indeed,
acting with the Hamiltonian one obtains
\begin{equation}
H|v\rangle = E_v |v\rangle, \ E_v = - \frac{J}{4} N,
\end{equation}
i.e.\ $|v\rangle$ is indeed an eigenstate, with the vacuum energy $E_v$ 
proportional to the number of lattice sites $N$. The total spin of the state
$|v\rangle$ is $S = N/2$. By acting with the lowering operator 
$S^- = \sum_x S_x^-$ on the vacuum state $|v\rangle$, one can construct the
other ground states belonging to the same $SU(2)_s$ multiplet, which contains
$2 S + 1 = N + 1$ degenerate states.

Let us now construct a somewhat unconventionally normalized hole state of
momentum $p$
\begin{equation}
|h p\rangle = \sum_x \exp(i p x) c_{x\up} |v\rangle = c_{p\up} |v\rangle.
\end{equation}
In order to check whether this is an eigenstate we compute
\begin{equation}
H |h p\rangle = \left([H,c_{p\up}] + c_{p\up} H\right) |v\rangle = 
\left(E_h(p) + E_v \right) |h p\rangle,
\end{equation}
which shows that $|h p\rangle$ is indeed an energy eigenstate. One obtains the
energy-momentum dispersion relation of a hole as
\begin{equation}
E_h(p) = \frac{J}{2} + \frac{U}{2} + 2 t \cos(p a) = 
\frac{J}{2} + \frac{U}{2} + 2 t - t a^2 \hat p^2,
\end{equation}
where we have introduced $\hat p = \frac{2}{a} \sin(p a/2)$. This 
periodic function has minima at $p = (2n-1) \pi/a$, with $n \in \mathbb{Z}$.
Expanding around $p = \pi/a$ we obtain
\begin{equation} \label{energyH}
E_h(p) = \frac{J}{2} + \frac{U}{2} - 2t +  t a^2 
\left(p - \frac{\pi}{a}\right)^2
+ {\cal O}\left(\left(p - \frac{\pi}{a}\right)^4\right).
\end{equation}
The holes are massive objects and their dispersion relation is given by
\begin{equation} 
\label{PH}
E_h(p) = M_h + \frac{(p - \pi/a)^2}{2M_h'}+{\cal O}\left(\left(p -
\frac{\pi}{a}\right)^4\right),
\end{equation}
with the rest mass $M_h$ and the kinetic mass $M_h'$ given by
\begin{equation}
M_h = \frac{J}{2} + \frac{U}{2} - 2 t, \quad M_h' = \frac{1}{2 t a^2}.
\end{equation}
Since the theory is non-relativistic, the rest mass $M_h$ and the kinetic mass
$M_h'$ need not to be the same.

Similarly, we construct electron states
\begin{equation}
|e p\rangle = \sum_x \exp(- i p x) c_{x\down} \Dag |v\rangle = 
c_{p\down} \Dag |v\rangle,
\end{equation}
and we compute
\begin{eqnarray}
Q^- |e p\rangle&=&\sum_x (-1)^{x/a} c_{x\down} c_{x\up} 
\sum_{x'} \exp(- i p x') c_{x'\down} \Dag |v\rangle \nonumber \\
&=&- \sum_x \exp\left(- i \left(p + \frac{\pi}{a}\right) x\right)
c_{x\up} |v\rangle = - |e -\left(p + \frac{\pi}{a}\right) \rangle.
\end{eqnarray}
The $SU(2)_{Q}$ symmetry then implies that the energy of an electron is given by
\begin{equation}
E_e(p) = \frac{J}{2} + \frac{U}{2} + 2t \cos(p a + \pi) = 
\frac{J}{2} + \frac{U}{2} - 2t + t a^2 \hat p^2.
\end{equation}
For electrons, the minima of the dispersion relation are located at $p = 2 n 
\pi/a$, with $n \in \mathbb{Z}$. Expanding around $p = 0$ we get
\begin{equation} \label{energyP}
E_e(p) = \frac{J}{2} + \frac{U}{2} - 2t + t a^2 p^2 + {\cal O}(p^4).
\end{equation}
Again, for small momenta
\begin{equation} 
E_e(p) = M_e + \frac{p^2}{2M'_e}.
\end{equation}
Due to the $SU(2)_Q$ symmetry, the rest and kinetic masses $M_{h,e}$ and
$M'_{h,e}$ of holes and electrons are identical.

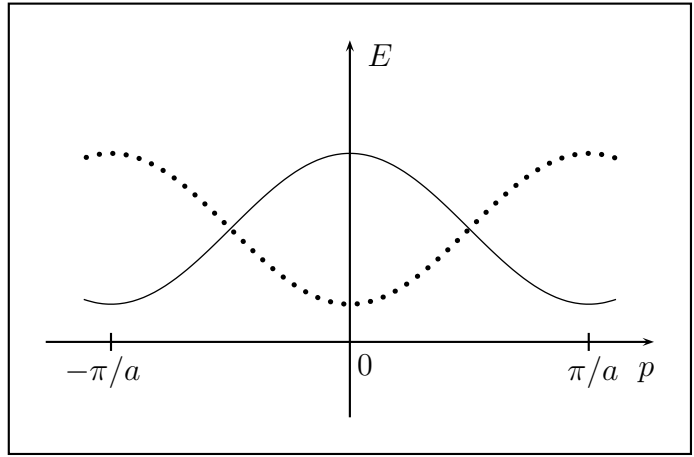
\begin{figure}[tbh]
\centering

\psset{unit=1cm} 
\begin{pspicture}(-4.5,-1.5)(4.5,4.5)
   \psclip{\psframe(-4.5,-1.5)(4.5,4.5)}
   \psaxes[Ox=0,Dx=3.141592654,Oy=0,Dy=6,labels=none]{->}(0,0)(-4,-1)(4,4)
   \psplot[plotstyle=curve,linestyle=dotted,linewidth=2pt]{-3.5}{3.5}%
   {x COS neg 1.5 add}%
   \psplot[plotstyle=curve,linewidth=0.5pt]{-3.5}{3.5}%
   {x COS 1.5 add}%
\rput(0.2,-0.3){$0$}
\rput(-3.25,-0.4){$-\pi/a$}
\rput(3.2,-0.4){$\pi/a$}
\rput(3.9,-0.4){$p$}  
  \rput(0.4,3.8){$E$}  
   \endpsclip
\end{pspicture}

  \caption{Dispersion relations for electrons (dotted curve) and holes (solid 
curve).} \label{notext}
\end{figure}

\subsection{Gap Equation for Magnon States}

As we have discussed before, in quantum ferromagnets the global spin rotational
symmetry $SU(2)_s$ is spontaneously broken by the formation of a uniform 
magnetization. The ground states of these systems are invariant only under
spin rotations in the subgroup $U(1)_s$. In this case, Goldstone's theorem
predicts $3 - 1 = 2$ massless boson fields --- the magnons --- also known as 
ferromagnetic spin waves.

A general ansatz for an electron-hole state is given by
\begin{equation}
|e h p\rangle = \sum_{x,y} \exp(i p y) f(x) c_{y\down} \Dag c_{y+x\up} |v\rangle.
\end{equation}
Here $x$ is the distance between the electron and the hole and $f(x)$ is the 
corresponding wave function of their relative motion. Indeed, magnons are 
massless bound states of an electron and a hole. We now consider
\begin{eqnarray}
H |e h p\rangle&=&H \sum_{x,y} \exp(i p y) f(x) c_{y\down} \Dag c_{y+x\up}  
|v\rangle \nonumber \\
&=&\left(\left[H,\sum_{x,y} \exp(i p y) f(x) c_{y\down} \Dag c_{y+x\up} \right] + 
\sum_{x,y} \exp(i p y) f(x) c_{y\down} \Dag c_{y+x\up} H \right) |v\rangle.
\nonumber \\ \
\end{eqnarray}
The last term on the right-hand side represents the vacuum energy. The
electron-hole energy $E_{eh}(p)$ is given by
\begin{equation} 
\label{spinwave1}
\left[H,\sum_{x,y} \exp(i p y) f(x) c_{y\down} \Dag c_{y+x\up}\right] |v\rangle =
E_{eh}(p) |e h p\rangle.
\end{equation}
A somewhat tedious evaluation of eq.(\ref{spinwave1}) implies that
$|e h p\rangle$ is an eigenstate only if
\begin{equation}
E_{eh}(p) f(x) = - t \left[f(x-a)(e^{ipa}-1)+f(x+a)(e^{-ipa}-1)\right] + 
(J + U)f(x),
\end{equation}
in the generic case $x \neq 0, \pm a$, as well as
\begin{equation}
E_{eh}(p) f(x) = - t \left[f(x-a)(e^{ipa}-1)+f(x+a)(e^{-ipa}-1)\right] + 
\left(\frac{3}{4} J + U\right) f(x), 
\end{equation}
in the special case $x = \pm a$, and
\begin{equation}
E_{eh}(p) f(x) = - t \left[f(x-a)(e^{ipa}-1)+f(x+a)(e^{-ipa}-1)\right] + 
\left(J \frac{a^2 \hat p^2}{2} + U\right) f(x),
\end{equation}
in the special case $x = 0$.
These three equations represent the lattice Schr\"odinger equation for an 
electron-hole pair with wave function $f(x)$. In order to solve these
equations, we transform to momentum space, i.e.
\begin{equation} 
f(q) = \sum_x f(x) \exp(- i q x), \quad 
f(x) = \frac{1}{2 \pi} \int_{-\pi/a}^{\pi/a} dq \ f(q) \exp(i q x),
\end{equation}
and we obtain
\begin{equation} 
\label{eq147}
f(q) = \frac{A \cos(q a) +  B \sin(q a) + C}{E_{eh}(p) - 2 t 
\left[\cos(q a) - \cos(q a - p a)\right] - J - U},
\end{equation}
with
\begin{eqnarray} \label{ABC1}
A&=&-\frac{J}{2} \frac{1}{2 \pi} \int_{-\pi/a}^{\pi/a} dq \ f(q) \cos(q a),
\nonumber \\
B&=&-\frac{J}{2} \frac{1}{2 \pi} \int_{-\pi/a}^{\pi/a} dq \ f(q) \sin(q a),
\nonumber \\
C&=&\left(J \frac{\hat p^2 a^2}{2} - J - U\right) \frac{1}{2 \pi}
\int_{-\pi/a}^{\pi/a} dq \ f(q) .
\end{eqnarray}
These are three coupled gap equations which must be solved self-consistently 
for $A, B, C$, and $E_{eh}(p)$. We are going to do this in the next subsection.
The denominator in eq.(\ref{eq147}) can be rewritten as
\begin{equation}
E_{eh}(p) - 2t \left[\cos(q a) - \cos(q a - p a)\right] - J - U =
\alpha \cos(q a) + \beta \sin(q a) + \gamma, 
\end{equation}
where
\begin{equation}
\alpha = 2 t [\cos(p a) - 1], \quad 
\beta = 2 t \sin(p a), \quad
\gamma = E_{eh}(p) - J - U.
\end{equation}

\subsection{Solution of the Gap Equation}

Let us now solve the gap equation (\ref{ABC1}). Inserting this equation into
eq.(\ref{ABC1}), we obtain an eigenvalue problem with eigenvalue 1
\begin{equation} 
\label{evprob}
I \left(\begin{array}{*{3}{c}} A \\ B \\ C \end{array} \right) =
- \frac{J}{2} \begin{pmatrix} I_4 & I_6 & I_2 \\ I_6 & I_5 & I_3 \\ 
z I_2 & z I_3 & z I_1  \end{pmatrix} 
\left(\begin{array}{*{3}{c}} A \\ B \\ C \end{array} \right) =
\left(\begin{array}{*{3}{c}} A \\ B \\ C \end{array} \right),
\end{equation}
where
\begin{eqnarray} 
\label{I1}
z&=&2 + 2 \frac{U}{J} - \hat p^2 a^2 \nonumber, \\
I_1&=&\frac{1}{2 \pi} \int_{-\pi/a}^{\pi/a} dq \ 
\frac{1}{\alpha \cos(q a) + \beta \sin(q a) + \gamma} =
 \mbox{\mbox{sign}}(\gamma) \frac{1}{s}, \nonumber \\
I_2&=&\frac{1}{2 \pi} \int_{-\pi/a}^{\pi/a} dq \ 
\frac{\cos(q a)}{\alpha \cos(q a) + \beta \sin(q a) + \gamma} = 
- \frac{\alpha}{s(|\gamma|+s)}, \nonumber \\
I_3&=&\frac{1}{2 \pi} \int_{-\pi/a}^{\pi/a} dq \ 
\frac{\sin(q a)}{\alpha \cos(q a) + \beta \sin(q a) + \gamma} =
- \frac{\beta}{s(|\gamma|+s)}, \nonumber \\
I_4&=&\frac{1}{2 \pi} \int_{-\pi/a}^{\pi/a} dq \ 
\frac{\cos^2(q a)}{\alpha \cos(q a) + \beta \sin(q a) + \gamma} = 
\mbox{sign}(\gamma) \frac{\alpha^{2}+s(|\gamma|+s)}{s(|\gamma|+s)^{2}}, 
\nonumber \\
I_5&=&\frac{1}{2 \pi} \int_{-\pi/a}^{\pi/a} dq \ 
\frac{\sin^2(q a)}{\alpha \cos(q a) + \beta \sin(q a) + \gamma} =
\mbox{sign}(\gamma) \frac{\beta^{2}+s(|\gamma|+s)}{s(|\gamma|+s)}, \nonumber \\
I_6&=&\frac{1}{2 \pi} \int_{-\pi/a}^{\pi/a} dq \ 
\frac{\cos(q a) \sin(q a)}{\alpha \cos(q a) + \beta \sin(q a) + \gamma} =
\mbox{sign}(\gamma) \frac{\alpha \beta}{s(|\gamma|+s)^{2}}, \nonumber \\
s&=&\sqrt{\gamma^2 - \alpha^2 - \beta^2}.
\end{eqnarray}
Using the values of these integrals which are considered in the appendix, the
eigenvalues of the matrix $I$ of eq.(\ref{evprob}) are given by
\begin{eqnarray} 
\label{i1}
i_1&=&- \mbox{sign}(\gamma)\frac{J}{2} \frac{1}{|\gamma|+s}, \nonumber \\
i_2&=&- \mbox{sign}(\gamma)\frac{J}{4} \frac{1}{s(|\gamma|+s)} \nonumber \\
&&\times \left[|\gamma|+z(|\gamma|+s)+\sqrt{-4s(|\gamma|+s)z+(|\gamma|+
z(|\gamma|+s))^2}\right], \nonumber \\
i_3&=&- \mbox{sign}(\gamma)\frac{J}{4} \frac{1}{s(|\gamma|+s)} \nonumber \\
&&\times \left[|\gamma|+z(|\gamma|+s)-\sqrt{-4s(|\gamma|+s)z+(|\gamma|+
z(|\gamma|+s))^2}\right].
\end{eqnarray}
The solutions of the gap equation (\ref{ABC1}) correspond to eigenvalues 1 in
eq.(\ref{evprob}). The condition $i_1 = 1$ can be fulfilled only for 
$\gamma < 0$ and then implies
\begin{equation} 
\label{i1sol}
E^{(1)}_{eh}(p) = \frac{3}{4} J + U - \frac{4 t^2 \hat p^2 a^2}{J}.
\end{equation}
Although this is the energy of an electron-hole state with total momentum $p$,
the corresponding eigenstate does not represent a magnon because
$E^{(1)}_{eh}(p)$ does not vanish for zero momentum. Similarly, the condition
$i_3 = 1$ can be fulfilled only for $\gamma < 0$ which then implies
\begin{equation} 
\label{hb}
E^{(2)}_{eh}(p) = - \frac{(4 J + 16 U) t^2}{J (3J + 4U)} p^2 a^2 + 
\frac{3}{4} J + U + {\cal O}(p^4).
\end{equation}
As before, this is indeed the energy of an electron-hole state, but this state
is not a magnon either. Finally, (again for $\gamma < 0$) the condition
$i_2 = 1$ implies
\begin{equation}
\label{hd}
E^{(3)}_{eh}(p) = \frac{J(3 J + 4 U) - 16 t^2}{2(3 J + 4 U)} p^2 a^2 + 
{\cal O}(p^4).
\end{equation}
This energy vanishes at zero momentum. Hence, the corresponding electron-hole
eigenstate can be identified as a magnon state. Indeed, the non-relativistic
dispersion relation $E^{(3)}_{eh}(p) \propto p^2$ is characteristic for 
ferromagnetic spin waves. In momentum space the wave function for the relative
motion of the electron and hole forming the massless magnon takes the form
\begin{equation}
f(q) = \frac{A \cos(q a) + B \sin(q a) + C}
{E^{(3)}_{eh}(p) - 2 t \left[\cos(q a) - \cos(q a - p a)\right] - J - U},
\end{equation}
which turns into
\begin{equation} 
\label{wfbeforetwo}
f(x) = {\cal N} \left(\frac{\alpha + i \beta}{|\gamma| + s}\right)^{x/a},
\end{equation}
for $x \geq 0$, where ${\cal N}$ is a normalization factor. For $x \leq 0$ one
finds $f(x) = f(-x)^*$.

\subsection{Two-Hole States}

Similar to the particle-hole spin-wave states, we now derive a Schr\"odinger 
equation for two-hole bound states. We make the ansatz
\begin{equation}
|hhp\rangle = \sum_{x,y} \exp(i p y)  g(x) c_{y\up} c_{y+x\up} |v\rangle, \quad 
g(- x)= - g(x) \exp(- i p x).
\end{equation}
The antisymmetry condition $g(- x)= - g(x) \exp(- i p x)$ follows from the
Pauli principle. In complete analogy to the particle-hole states, one derives
the Schr\"odinger equation
\begin{equation} 
\label{schroed_th}
E_{hh}(p) g(x) = t \left[g(x+a) (1 + e^{-i p a}) + g(x-a) (1 + e^{i p a})\right]
+ (J + U)g(x),
\end{equation}
for a generic situation with $x \neq 0,\pm a$. In the special case $x = \pm a$
one obtains
\begin{equation}
E_{hh}(p) g(x) = t \left[g(x+a) (1 + e^{-i p a}) + g(x-a) (1 + e^{i p a})\right]
+ \left(\frac{3}{4} J + U\right)g(x),
\end{equation}
while for $x = 0$ the Schr\"odinger equation takes the form
\begin{equation}
E_{hh}(p) g(x) = t \left[g(x+a) (1 + e^{-i p a}) + g(x-a) (1 + e^{i p a})\right].
\end{equation}
Going to momentum space
\begin{equation}
g(q) = \sum_x g(x) \exp(- i q x), \quad 
g(x) = \frac{1}{2 \pi} \int_{-\pi/a}^{\pi/a} dq \ g(q) \exp(i q x),
\end{equation}
one obtains the gap equation
\begin{eqnarray}
A&=&-\frac{J}{2} \frac{1}{2 \pi} \int_{-\pi/a}^{\pi/a} dq \ g(q) \cos(q a),
\nonumber \\
B&=&-\frac{J}{2} \frac{1}{2 \pi} \int_{-\pi/a}^{\pi/a} dq \ g(q) \sin(q a).
\end{eqnarray}
with
\begin{equation}
g(q) = \frac{A \cos(q a) +  B \sin(q a)}
{E_{hh}(p) - 2 t \left[\cos(q a) - \cos(q a - p a)\right] - J - U},
\end{equation}
The antisymmetry condition $g(- x)= - g(x) \exp(- i p x)$ implies
$g(- q) = - g(p+q)$. Imposing this condition on the gap equation leads to
\begin{equation}
\begin{pmatrix} - \cos(p a) & - \sin(p a) \\ - \sin(p a) & \cos(p a) 
\end{pmatrix} \left( \begin{array}{*{2}{c}} A \\ B \\ \end{array}\right) = 
\left( \begin{array}{*{2}{c}} A \\ B \\ \end{array}\right) \Rightarrow 
\cos\left(\frac{p a}{2}\right) A = - \sin\left(\frac{p a}{2}\right) B .
\end{equation}
We now introduce $C$, such that
\begin{equation}
A = C \sin\left(\frac{p a}{2}\right), \quad 
B = - C \cos\left(\frac{p a}{2}\right).
\end{equation}
The gap equation can thus be written as

\begin{equation} \label{gaptwoc}
C = \frac{J}{2} \frac{1}{2 \pi} \int_{-\pi/a}^{\pi/a} dq \ g(q) 
\sin\left(q a - \frac{p a}{2}\right).
\end{equation}
with
\begin{equation} 
\label{gaptwo}
g(q) = \frac{- C \sin(q a - p a/2)}
{E_{hh}(p) - 2 t \left[\cos(p a - q a) + \cos(q a)\right] - J - U},
\end{equation}

Let us now solve the gap equation. Inserting eq.(\ref{gaptwo}) into 
eq.(\ref{gaptwoc}) yields
\begin{equation}
1 = - \frac{J}{2} \frac{1}{2 \pi} \int_{-\pi/a}^{\pi/a} dq \ \frac{\sin^2(q a)}
{E_{hh}(p) - 4 t \cos(q a) \cos(p a/2) - J - U}.
\end{equation}
Using
\begin{equation}
\frac{1}{2 \pi} \int_{-\pi/a}^{\pi/a} dq \ \frac{\sin^2(q a)}
{\alpha \cos(q a) + \gamma} = \mbox{sign}(\gamma) 
\frac{1}{|\gamma| + \sqrt{\gamma^2 - \alpha^2}},
\end{equation}
with
\begin{equation} \label{ca}
\gamma = E_{hh}(p) - J - U, \quad \alpha = -4 t \cos\left(\frac{p a}{2}\right),
\end{equation}
for $\gamma < 0$ one thus obtains
\begin{equation}
\frac{J}{2} = J + U - E_{hh}(p) + 
\sqrt{(E_{hh}(p) - J - U)^2 - 16 t^2 \cos^2(p a/2)}.
\end{equation}
Squaring this equation we find the two-hole energy
\begin{equation} \label{htwo}
E_{hh}(p) = \frac{3}{4} J + U - \frac{16t^2}{J} + \frac{4 t^2 p^2 a^2}{J} + 
{\cal O}(p^4).
\end{equation}
From this expression we read off the total rest mass of the two-hole bound 
state as 
\begin{equation}
M_{hh} = \frac{3}{4} J + U - \frac{16t^2}{J},
\end{equation}
while the corresponding kinetic mass is given by
\begin{equation}
\label{mkinhh}
M_{hh}' = \frac{J}{8 t^2 a^2}.
\end{equation}
The binding energy of the two-hole state hence takes the form
\begin{equation} \label{Eb_mmf}
E_B = 2 M_h - M_{hh} = J \left(\frac{1}{2} - \frac{4t}{J}\right)^2.
\end{equation}
As we will see below, the two-hole bound-state wave function is normalizable 
only for $0 < t < J/8$, while for $t > J/8$ the two-hole bound state
disappears.

Let us now consider the wave function in coordinate space. For $\gamma < 0$
we obtain
\begin{equation} 
\label{wfcoord}
g(x) = C_0 \exp\left( \frac{i p x}{2}\right)
\left(- 8 \frac{t}{J} \cos\left(\frac{p a}{2}\right)\right)^{x/a},
\end{equation}
where $C_0$ is a normalization constant. The expression eq.(\ref{wfcoord}) 
indeed satisfies the antisymmetry condition $g(- x) = - g(x) \exp(- i p x)$.
%
%
%

Let us also consider scattering states of two holes described by the ansatz
\begin{equation} 
\label{scattering_MMF}
g(x) = \widetilde A \exp(i q x) + \widetilde B \exp(- i q x).
\end{equation}
For $x \neq 0,\pm a$ and $p=0$ the Schr\"odinger equation is given by
\begin{equation} 
\label{sr1}
2t \left[g(x+a)+g(x-a)\right] + (J + U) g(x) = E g(x).
\end{equation}
Inserting the ansatz of eq.(\ref{scattering_MMF}) into eq.(\ref{sr1}) leads to
\begin{equation} 
\label{scenrg}
E = 4 t \cos(q a) + J + U,
\end{equation}
while the amplitudes are given by
\begin{equation}
\widetilde A = \frac{J}{16t} + i \frac{J\cos(q a) + 8t}{16 t \sin(q a)}, \quad 
\widetilde B = \frac{J}{16t} - i \frac{J\cos(q a) + 8t}{16 t \sin(q a)}
= \widetilde A^*.
\end{equation}
This implies that the ansatz of eq.(\ref{scattering_MMF}) is purely real.
Finally, we obtain
\begin{equation} 
\label{relMMF}
\frac{\widetilde A}{\widetilde B} = 
\frac{\sin(q a) + i (\cos(q a) + 8t/J)}
{\sin(q a) - i (\cos(q a) + 8t/J)}.
\end{equation}
Later we will compare this result with the corresponding one obtained in the 
effective field theory.

\section{Construction of the Effective Field Theory for the Doped
Ferromagnet}

Before we go into details, we would like to make some general remarks about the
next subsections. In section 2, we discussed a microscopic model describing a
doped ferromagnet. With this model we were able to calculate dispersion 
relations for magnon-, electron-, and hole-states. The effective field theory 
we are going to discuss now captures the low-energy physics of the underlying
microscopic system, order by order in a systematic low-energy expansion. The 
effective field theory is constructed in complete analogy to the corresponding
cases of hole- or electron-doped antiferromagnets \cite{Kae05,Bru06,Bru07}.
The antiferromagnetic systems are of particular physical interest due to their 
relation with high-temperature superconductors. In contrast to the
ferromagnetic case discussed here, the microscopic Hubbard-type models for
doped antiferromagnets can not be solved analytically. Hence, in that case,
one must rely on numerical simulations in order to test the low-energy
effective theory and to fix its low-energy parameters. The ferromagnetic
system studied in this paper, on the other hand, provides an exceptional case
in which the predictions of the effective theory can be tested against exact
analytic results in the underlying microscopic model. By comparing results in
the magnon, single-hole, as well as two-hole sectors, we will be able to fix
the a priori undetermined low-energy parameters of the effective theory and
even test it beyond perturbation theory. The quantitative agreement that
is achieved in the ferromagnetic case lends further support to the effective
field theory approach also for the physically most relevant antiferromagnets,
since the basic principles underlying both constructions are identical.

\subsection{Symmetry Properties of Magnon Fields}

In this subsection, we are going to investigate the symmetries of magnon
fields. At the beginning of section 2, we have studied the symmetries of the 
microscopic model describing ferromagnetism. The effective field theory must
share the symmetries of the underlying microscopic system. Therefore we now
construct magnon fields and discuss how they transform under those symmetries.

As we have mentioned in section 2, in a quantum ferromagnet the global spin 
rotation symmetry $G = SU(2)_s$ is spontaneously broken by the formation of a 
uniform magnetization. The ground state of these systems is invariant only 
under spin rotations in the unbroken subgroup $H = U(1)_s$. As a consequence of
the spontaneous symmetry breaking, there are two massless Goldstone boson 
fields, leading to the ferromagnetic spin wave or magnon. We already discussed
magnons in the microscopic model, where we calculated the dispersion relation
in eq.(\ref{hd}). In the effective field theory the direction of the
magnetization is described by a unit-vector field
\begin{equation} 
\label{magnon1}
\vec e(x) = (e_1(x),e_2(x),e_3(x)) \in S^2, \quad \vec e(x)^2 = 1,
\end{equation}
in the coset space $G/H = SU(2)_s/U(1)_s = S^2$, where $x = (x_1,t)$
is a point in $(1+1)$-dimensional Euclidean space-time.

Beyond the $O(3)$ vector representation $\vec e(x)$, it is useful to introduce
an alternative $\mathbb{C}P(1)$ representation of the magnon field using
$2 \times 2$ Hermitean  projection matrices $P(x)$ that obey
\begin{eqnarray}
P(x)\Dag = P(x), \quad \Tr P(x) = 1, \quad P(x)^2 = P(x),
\end{eqnarray}
and are given by
\begin{equation} 
\label{magnon2}
P(x) = \frac{1}{2}(\mathds{1} + \vec{e}(x) \cdot \vec \sigma) = \frac{1}{2}
\begin{pmatrix} 1 + e_3(x) & e_1(x) - i e_2(x) \\ e_1(x) + i e_2(x) & 1 - e_3(x)
\end{pmatrix}.
\end{equation}

The first symmetry we encountered in section 2 was the global spin rotation 
symmetry $SU(2)_s$ under which the magnon field transforms as
\begin{equation} 
P(x)' = g P(x) g\Dag. 
\end{equation}
Note that the magnon field $P(x)$ is invariant under the Abelian and 
non-Abelian fermion number symmetries $U(1)_Q$ and $SU(2)_Q$, i.e.\ 
\begin{equation}
^{\vec Q} P(x) = P(x).
\end{equation}
Unlike in an antiferromagnet, in a ferromagnet the order parameter 
$\vec e(x)$ is invariant under the displacement symmetry $D$, i.e. 
\begin{eqnarray} 
^D\vec e(x) = \vec e(x) \ \Rightarrow \ ^D P(x) = P(x).
\end{eqnarray}
Under the spatial reflection $R$ which turns the point $x = (x_1,t)$
into the reflected point $Rx = (-x_1,t)$ the magnon field transforms as
\begin{eqnarray} 
^R P(x) = P(Rx).
\end{eqnarray}
Another important symmetry is time reversal $T$ which turns $x$ into 
$Tx = (x_1,- t)$. The 
spin transforms like the orbital angular momentum $\vec L$ of a particle. The 
momentum $\vec p$ changes sign under time reversal and so does $\vec L$, i.e.\
$^T \vec L = - \vec L$. Consequently, under $T$ the magnetization vector (which
is a sum of microscopic spins) transforms as
\begin{eqnarray} 
^T\vec e(x) = - \vec e(Tx) \ \Rightarrow \ ^T P(x) = \mathds{1} - P(Tx).
\end{eqnarray}

\subsection{Effective Action for Magnons}

Since the low-energy physics is dominated by terms with the smallest possible 
number of derivatives, we construct an effective Lagrangian according to a 
systematic derivative expansion. All terms in the Lagrangian must be invariant 
under the symmetry transformations considered in the previous subsection. In
contrast to an antiferromagnet, a ferromagnet has a conserved order parameter
--- the total spin. In the effective theory, this manifests itself by the
presence of a Wess-Zumino term, which gives rise to a non-relativistic magnon
dispersion relation \cite{Leu94}. Indeed, as we have seen from the calculations
in the microscopic model, the ferromagnet has a non-relativistic spectrum, 
i.e.\ $E \sim p^2$. The leading order Euclidean effective action for an undoped
ferromagnet derived in \cite{Leu94} takes the form
\begin{equation} 
\label{actionmagnon}
S[\vec e] = \int dx_1 \int_0^\beta dt \ \frac{\rho_s}{2} 
\partial_1 \vec e \cdot \partial_1 \vec e  + S_{WZ}[\vec e],
\end{equation}
with $\rho_s$ being the spin stiffness. The Wess-Zumino term is given by
\begin{equation} 
\label{WESS}
S_{WZ}[\vec e] = - i m \int dx_1 \int_0^\beta dt \int_0^1 d\tau \ 
\vec e \cdot (\partial_t \vec e \times \partial_\tau \vec e).
\end{equation}
Here $m$ is the magnetization density. This term contains only one temporal 
derivative and hence leads to a non-relativistic dispersion relation. The 
coordinates $t$ and $\tau$ parameterize a disc or two-dimensional hemisphere
$H^2$, which is bounded by the compactified Euclidean time interval $S^1$. The 
magnon field $\vec e(x)$ at physical space-time points $x \in \mathbb{R} 
\times S^1$ is extended to a field $\vec e(x_1,t,\tau)$ in the 3-dimensional
domain $(x_1,t,\tau) \in \mathbb{R} \times H^2$. The integrand of the
Wess-Zumino term is a total derivative and hence only receives contributions
from the boundary, which coincides with the physical space-time where
$\vec e(x,\tau = 1) = \vec e(x)$. A possible extrapolation of the physical
magnon field into the additional dimension with $\vec e(x,\tau = 0) = (0,0,1)$
is given by
\begin{equation}
e_1(x,\tau) = \tau e_1(x), \quad
e_2(x,\tau) = \tau e_2(x), \quad
e_3(x,\tau) = \sqrt{1 - e _1(x,\tau)^2 - e_2(x,\tau)^2}.
\end{equation}

The action of eq.(\ref{actionmagnon}) enters the Euclidean path integral
\begin{equation} \label{path}
Z = \int \mathcal{D}\vec e \ \exp(- S[\vec e]),
\end{equation}
which should depend only on the physical magnon field and not on a particular 
extrapolation into the additional dimension. In order to show that this is 
indeed the case, we compare two arbitrary extrapolations
$\vec e^{\, (1)}(x,\tau)$ and $\vec e^{\, (2)}(x,\tau)$ and we consider the
difference between the two corresponding Wess-Zumino terms
\begin{eqnarray}
S_{WZ}[\vec e^{\,(1)}] - S_{WZ}[\vec e^{\,(2)}]&=&- i m \int dx_1 \int_{H^2} dt
d\tau \  \vec e^{\,(1)} \cdot (\partial_t \vec e^{\,(1)} \times \partial_\tau
\vec e^{\,(1)}) \nonumber \\
& &+ i m \int dx_1 \int_{H^2} dt d\tau \ \vec e^{\,(2)} \cdot (\partial_t
\vec e^{\,(2)} \times \partial_\tau \vec e^{\,(2)}) \nonumber \\
&=&- i m \int dx_1 \int_{S^2} dt d\tau \ \vec e \cdot
(\partial_t \vec e \times \partial_\tau \vec e).
\end{eqnarray}
The two extrapolations $\vec e^{\,(1)}$ and $\vec e^{\,(2)}$ over the two
hemispheres $H^2$ (which are differently oriented due to the minus sign
between the Wess-Zumino terms) are combined to an extrapolation $\vec e$ over
an entire compact sphere $S^2$. We now use the fact that
\begin{equation}
\frac{1}{4\pi} \int_{S^2} dt d\tau \ \vec e \cdot (\partial_t \vec e \times 
\partial_\tau \vec e) = n \in \mathbb{Z}
\end{equation}
is the integer winding number of the field $\vec e$ which maps $S^2$ 
(parameterized by $t$ and $\tau$) into the order parameter sphere $S^2$. Indeed,
the corresponding second homotopy group is given by $\Pi_2[S^2] = \mathbb{Z}$.
Hence, the extrapolation ambiguity is given by
\begin{eqnarray}
S_{WZ}[\vec e^{\,(1)}] - S_{WZ}[\vec e^{\,(2)}] = - i m \int dx_1 \ 4 \pi n.
\end{eqnarray}
Since $m$ is the magnetization density,
\begin{eqnarray} \label{mdensity}
M = m \int dx_1
\end{eqnarray}
is the total spin of the entire magnet and hence an integer or a half-integer. 
Since
\begin{eqnarray}
\exp(- S_{WZ}[\vec e^{\,(1)}] + S_{WZ}[\vec e^{\,(2)}]) = \exp(4 \pi i M n)
= 1,
\end{eqnarray}
the factor $\exp(- S[\vec e])$ that enters the path integral of eq.(\ref{path})
is thus unambiguously defined, irrespective of the arbitrarily chosen
extrapolation ${\vec e}(x,\tau)$.

In the $\mathbb{C}P(1)$ representation, the leading order low-energy
Euclidean action takes the form
\begin{equation} 
\label{action1}
S[P] = \int dx_1 \int_0^\beta dt \ \Tr \left[\rho_s \partial_1 P\partial_1 P +
2 m \int_0^1 d\tau \ P(\partial_t P \partial_\tau P - \partial_\tau P
\partial_t P) \right].
\end{equation}
From eq.(\ref{actionmagnon}) one can derive the Landau-Lifshitz equation for 
spin waves in a ferromagnet \cite{Lan81}
\begin{equation}
\rho_s \vec e \times \partial_1^{2} \vec e = m \partial_t \vec e.
\end{equation}
We assume a magnetization in the $3$-direction with small perturbations in the 
$1$- and $2$-directions, i.e.
\begin{equation}
\vec e(x) = \left(\frac{m_1(x)}{\sqrt{\rho_s}},\frac{m_2(x)}{\sqrt{\rho_s}},1
\right) + {\cal O}(m^2).
\end{equation}
Expanding up to linear powers in the magnon fluctuations $m_1$ and $m_2$ one 
obtains the equation
\begin{equation}
i \partial_t (m_1 + i m_2) = - \frac{\rho_s}{m} \partial_1^2 (m_1 + i m_2),
\end{equation}
which implies the non-relativistic magnon dispersion relation
\begin{eqnarray} 
\label{spinwaveeft}
E_m(p) = \frac{\rho_s p^2}{m}.
\end{eqnarray}

\subsection{Determination of the Low-Energy Parameters}

At this point, we can match the low-energy parameters $m$ and $\rho_s$ of the 
effective field theory to the coupling constants $t$, $J$, and $U$ of the
underlying microscopic system. First of all, in the ferromagnetic ground state
all spins are up, and hence the magnetization density is given by
\begin{equation}
m = \frac{1}{2 a}.
\end{equation}
The magnon dispersion relation obtained in the microscopic model was given by
\begin{equation}
E^{(3)}_{eh}(p) = \frac{J(3J + 4U) - 16 t^2}{2(3 J + 4 U)} p^2 a^2+ {\cal O}
\left(p^4\right).
\end{equation}
Identifying $E^{(3)}_{eh}(p)$ with $E_m(p)$ we read off the value
\begin{equation}
\rho_s = \frac{J(3J + 4U) - 16 t^2}{4(3 J + 4 U)} a,
\end{equation}
for the spin stiffness. For $t = 0$ the microscopic model reduces to the 
Heisenberg model and the spin stiffness takes the familiar value
$\rho_s = J a /4$. It should be noted that the ferromagnetic vacuum
becomes unstable when $16 t^2 > J(3J + 4U)$. 

\subsection{Non-linear Realization of the $SU(2)_s$ Spin Symmetry}

In order to couple electron or hole fields to the order parameter, a non-linear
realization of the $SU(2)_s$ symmetry has been constructed in
\cite{Kae05,Bru06,Bru07}. A
local transformation $h(x)\in U(1)_s$ is then constructed from the global
transformation $g \in SU(2)_s$ as well as from the local magnon field $P(x)$
as follows. First, one diagonalizes the magnon field by a unitary
transformation  $u(x) \in SU(2)_s$, i.e. 
\begin{equation}
u(x) P(x) u(x)\Dag = \frac{1}{2} (\mathds{1} + \sigma_3) = 
\begin{pmatrix} 1 & 0 \\ 0 & 0 \end{pmatrix}, \quad u_{11}(x) \geq 0.
\end{equation}
Note that, due to its projector properties, $P(x)$ has eigenvalues 0 and 1. In 
order to make $u(x)$ uniquely defined, we demand that the element $u_{11}(x)$ 
is real and non-negative. Otherwise, the diagonalizing matrix $u(x)$ would be 
defined only up to a $U(1)_s$ phase. Using eq.(\ref{magnon2}) and spherical 
coordinates for $\vec e(x)$, i.e.
\begin{equation}
\vec e(x) = (\sin\theta(x) \cos\varphi(x), \sin\theta(x) \sin\varphi(x), 
\cos\theta(x)),
\end{equation}
one obtains
\begin{eqnarray}
u(x)&=&\frac{1}{\sqrt{2(1 + e_3(x))}}
\begin{pmatrix}
1 + e_3(x) & e_1(x) - i e_2(x) \\ - e_1(x) - i e_2(x) & 1 + e_3(x) \end{pmatrix}
\nonumber \\
&=&\begin{pmatrix}
\cos(\theta(x)/2) & \sin(\theta(x)/2) \exp(- i \varphi(x)) \\
- \sin(\theta(x)/2) \exp(i \varphi(x)) & \cos(\theta(x)/2) \end{pmatrix}.
\end{eqnarray}
Under a global $SU(2)_s$ transformation $g$ the diagonalizing field $u(x)$ 
transforms as
\begin{equation} \label{u_transform}
u(x)' = h(x) u(x) g\Dag, \quad u_{11}(x)' \geq 0,
\end{equation}
which implicitly defines the non-linear symmetry transformation
\begin{equation} 
\label{h}
h(x) = \exp(i \alpha(x) \sigma_3) = \begin{pmatrix}
\exp(i \alpha(x)) & 0 \\ 0 & \exp(- i \alpha(x)) \end{pmatrix} \in U(1)_s.
\end{equation}
The transformation $h(x)$ is uniquely defined since we demand that $u_{11}(x)'$
is again real and non-negative.

Since in a ferromagnet the order parameter is invariant under the displacement 
symmetry $D$, we have
\begin{equation}
^D u(x) = u(x).
\end{equation}
In order to couple electrons and holes to the magnons it is necessary to 
introduce the anti-Hermitean traceless field
\begin{equation}
v_\mu(x) = u(x) \partial_\mu u(x)\Dag,
\end{equation}
which under $SU(2)_s$ transforms as
\begin{equation}
v_\mu(x)' = h(x) u(x) g\Dag \partial_\mu \left[g u(x)\Dag h(x)\Dag \right] = 
h(x) \left[v_\mu(x) + \partial_\mu\right] h(x)\Dag.
\end{equation}
Since the field $v_\mu(x)$ is traceless, it can be written as a linear 
combination of the Pauli matrices $\sigma_a$
\begin{equation}
v_\mu(x) = i v_\mu^a(x) \sigma_a, \quad a \in \left\lbrace 1,2,3 \right\rbrace,
\quad v_\mu^a(x) \in \mathbb{R}.
\end{equation}
The factor $i$ is needed to make $v_\mu(x)$ anti-Hermitean. Introducing
\begin{equation} 
\label{vpm}
v_\mu^\pm(x) = v_\mu^1(x) \mp i v_\mu^2(x),
\end{equation}
we write
\begin{equation}
v_\mu(x) = i \begin{pmatrix} v_\mu^3(x) &  v_\mu^+(x) \\
v_\mu^-(x) & - v_\mu^3(x) \end{pmatrix}.
\end{equation}
This leads to the transformation laws for $v_\mu^3(x)$
\begin{align} 
\label{vmutrans}
SU(2)_s: & \quad v_\mu^3(x)' = v_\mu^3(x) - \partial_\mu \alpha(x), \nonumber \\
U(1)_Q: & \quad ^Q v_\mu^3(x) = v_\mu^3(x), \nonumber \\
D: & \quad ^D v_\mu^3(x) = v_\mu^3(x), \nonumber \\
R: & \quad ^R v_1^3(x) = - v_1^3(Rx), \quad ^R v_t^3(x) = v_t^3(Rx),
\nonumber \\
T: & \quad ^T v_1^3(x) = - v_1^3(Tx), \quad ^T v_t^3(x) = v_t^3(Tx),
\end{align}
as well as for $v_\mu^\pm(x)$
\begin{align} 
\label{vmupmt}
SU(2)_s: & \quad v_\mu^\pm(x)' = \exp(\pm 2 i \alpha(x)) v_\mu^\pm(x), 
\nonumber \\
U(1)_Q: & \quad ^Q v_\mu^\pm(x) = v_\mu^\pm(x), \nonumber \\
D: & \quad ^D v_\mu^\pm(x) = v_\mu^\pm(x), \nonumber \\
R: & \quad ^R v_1^\pm(x) = - v_1^\pm(Rx), \quad ^R v_t^\pm(x) = v_t^\pm(Rx),
\nonumber \\
T: & \quad ^T v_1^\pm(x) = - v_1^\pm(Tx), \quad ^T v_t^\pm(x) = v_t^\pm(Tx).
\end{align}

\subsection{Microscopic Operators in a Magnon Background Field}

In the context of effective field theory, until now we have only discussed 
magnons, which correspond to states at half-filling in the microscopic model. 
In this subsection, we will begin to include doped electrons and holes. For
this purpose, we must establish a connection between the microscopic degrees of
freedom and the low-energy effective fields describing electrons or holes. 
Following \cite{Kae05,Bru06,Bru07}, we now discuss how this connection is
established. It is a virtue of the completely analytically controlled
ferromagnetic case that this connection can be tested rigorously.

As discussed in detail in \cite{Kae05,Bru06,Bru07}, in order to define new
operators $\Psi_x^A$ and $\Psi_x^B$ it is useful to introduce the matrix-valued
fermion operator $C_x$. We have already used the operator $C_x$ in
eq.(\ref{H_nice}) to rewrite the Hamiltonian in a manifestly
$SU(2)_s$-, $SU(2)_Q$-, $D$-, and $R$-invariant form. Now we write
\begin{eqnarray}
&&\Psi_x^A = u(x) C_x = u(x) \begin{pmatrix} c_{x\up} & c_{x\down}\Dag \\ 
c_{x\down} & - c_{x\up}\Dag \end{pmatrix} = 
\begin{pmatrix} \psi_{x+}^A & \psi_{x-}^{A\dagger} \\ \psi_{x-}^A & 
- \psi_{x+}^{A\dagger} \end{pmatrix}, \quad x \in A, \nonumber \\
&&\Psi_x^B = u(x) C_x = u(x) \begin{pmatrix} c_{x\up} & - c_{x\down}\Dag \\ 
c_{x\down} & c_{x\up}\Dag \end{pmatrix} = 
\begin{pmatrix} \psi_{x+}^B & -\psi_{x-}^{B\dagger} \\ \psi_{x-}^B & 
\psi_{x+}^{B\dagger} \end{pmatrix}, \quad  x \in B.
\end{eqnarray}
Note that $\Psi$ denotes a matrix while $\psi$ denotes a matrix element. The 
new lattice operators inherit their transformation properties from the 
operators of the microscopic model, i.e.\ we use the transformation properties 
of $C_x$ discussed in section 2. It should be noted that here the continuum 
field $u(x)$ is evaluated only at discrete lattice points $x$. According to 
eq.(\ref{F_transform}) and eq.(\ref{u_transform}), under the $SU(2)_s$ symmetry
one obtains
\begin{equation}
\Psi_x^{A,B'} = u(x)' C_x' = h(x) u(x) g\Dag g C_x = h(x) \Psi_x^{A,B}.
\end{equation}
In components this relation takes the form
\begin{equation}
\psi_{x\pm}^{A,B'} = \exp(\pm i\alpha(x)) \psi_{x\pm}^{A,B}.
\end{equation}
Similarly, under the $SU(2)_Q$ symmetry one obtains
\begin{equation}
^{\vec Q} \Psi_x^{A,B} = ^{\vec Q} \!\! u(x) ^{\vec Q} C_x = u(x) C_x
\Omega^T = \Psi_x^{A,B} \Omega^T.
\end{equation}
Here we have used the fact that $u(x)$ is invariant under the fermion number 
symmetries $U(1)_Q$ and $SU(2)_Q$, i.e.\ $^{\vec Q} u(x) = u(x)$. In particular,
under the $U(1)_Q$ subgroup of $SU(2)_Q$ the components transform as
\begin{equation}
^Q \psi_{x\pm}^{A,B} = \exp(i \omega) \psi_{x\pm}^{A,B}.
\end{equation}
Under the displacement symmetry we obtain
\begin{equation}
^D \Psi_x^{A,B} = ^D \!\! u(x) ^D C_x^{A,B} = 
u(x + a) C_{x + a}^{B,A} \sigma_3 = \Psi_{x + a}^{B,A} \sigma_3.
\end{equation}
Expressed in terms of components this implies
\begin{equation}
^D \psi_{x\pm}^{A,B} = \psi_{x + a \pm}^{B,A}.
\end{equation}

\subsection{Effective Fields for Charge Carriers}

In the low-energy effective field theory we will use a Euclidean path integral 
description instead of the Hamiltonian description used in the microscopic 
model. The lattice operators $\psi_{x\pm}^{A,B}$ and
$\psi_{x\pm}^{A,B\dagger}$ are then replaced by Grassmann numbers
$\psi^{A,B}_{\pm}(x)$ and $\psi^{A,B\dagger}_{\pm}(x)$ which are completely
independent of each other. Therefore, in the effective field theory the
electron and hole fields are represented by eight independent Grassmann
numbers $\psi^{A,B}_{\pm}(x)$ and 
$\psi^{A,B\dagger}_{\pm}(x)$ which can be combined to
\begin{equation}
\Psi^A(x) = \begin{pmatrix}
\psi^A_+(x) & \psi^{A\dagger}_-(x) \\ \psi^A_-(x) & - \psi^{A\dagger}_+(x)
\end{pmatrix}, \quad 
\Psi^B(x) = \begin{pmatrix}
\psi^B_+(x) & - \psi^{B\dagger}_-(x) \\ \psi^B_-(x) & \psi^{B\dagger}_+(x)
\end{pmatrix}.
\end{equation}
For notational convenience we also introduce the fields
\begin{equation}
\Psi^{A\dagger}(x) = \begin{pmatrix}
\psi^{A\dagger}_+(x) &  \psi^{A\dagger}_-(x) \\ \psi^A_-(x) & - \psi^A_+(x)
\end{pmatrix}, \quad 
\Psi^{B\dagger}(x) = \begin{pmatrix}
\psi^{B\dagger}_+(x) & \psi^{B\dagger}_-(x) \\ - \psi^B_-(x) & \psi^B_+(x)
\end{pmatrix}.
\end{equation}
We should note that $\Psi^{A,B\dagger}(x)$ is not independent of
$\Psi^{A,B}(x)$, since both contain the same Grassmann fields
$\psi^{A,B}_\pm(x)$ and $\psi^{A,B\dagger}_\pm(x)$. It should also be pointed
out that the continuum fields of the low-energy effective theory cannot be
derived explicitly from the lattice operators of the microscopic model. Still,
the Grassmann fields $\Psi^{A,B}(x)$ describing electrons and holes in the
low-energy effective theory transform just like the lattice operators
$\Psi_x^{A,B}$ discussed before. In contrast to the lattice operators, the
fields $\Psi^{A,B}(x)$ are defined in the continuum. Hence, under the
displacement symmetry $D$ one no longer distinguishes between the points $x$
and $x + a$.

We now list the transformation properties of the effective fields under the 
various symmetries, which can be derived using the transformation properties 
discussed above
\begin{align}
\label{transfderivations}
SU(2)_s: & \quad \Psi^{A,B}(x)' = h(x) \Psi^{A,B}(x), \quad 
\Psi^{A,B\dagger}(x)' = \Psi^{A,B\dagger}(x) h(x)\Dag, \nonumber \\
SU(2)_Q: & \quad ^{\vec Q}\Psi^{A,B}(x) = \Psi^{A,B}(x) \Omega^T, \quad 
^{\vec Q} \Psi^{A,B\dagger}(x) = \Omega^{T\Dag} \Psi^{A,B\dagger}(x), \nonumber \\
D: & \quad ^D \Psi^{A,B}(x) = \Psi^{B,A}(x) \sigma_3, \quad 
^D \Psi^{A,B\dagger}(x) = \sigma_3 \Psi^{B,A\dagger}(x), \nonumber \\
R: & \quad ^R \Psi^{A,B}(x) = \Psi^{A,B}(Rx), \quad 
^R\Psi^{A,B\dagger}(x) = \Psi^{A,B\dagger}(Rx), \nonumber \\
T: & \quad ^T \Psi^{A,B}(x) = - [\Psi^{A,B\dagger}(Tx)^T] \sigma_3, \quad 
^T \Psi^{A,B\dagger}(x) = \sigma_3 [\Psi^{A,B}(Tx)^T].
\end{align}
Note, that an upper index $T$ on the right denotes transpose, while on the
left it denotes time reversal. In components the symmetry transformations read
\begin{align}
SU(2)_s: & \quad \psi^{A,B}_\pm(x)' = \exp(\pm i\alpha(x)) \psi^{A,B}_\pm(x), 
\quad \psi^{A,B\dagger}_\pm(x)' = \exp(\mp i\alpha(x)) \psi^{A,B\dagger}_\pm(x), 
\nonumber \\
U(1)_Q: & \quad ^Q \psi^{A,B}_\pm(x) = \exp(i \omega) \psi^{A,B}_\pm(x), \quad 
^Q \psi^{A,B\dagger}_\pm(x) = \exp(- i \omega) \psi^{A,B\dagger}_\pm(x),
\nonumber \\
D: & \quad ^D \psi^{A,B}_\pm(x) = \psi^{B,A}_\pm(x), \quad 
^D \psi^{A,B\dagger}_\pm(x) = \psi^{B,A\dagger}_\pm(x), \nonumber \\
R: & \quad ^R \psi^{A,B}_\pm(x) = \psi^{A,B}_\pm(Rx), \quad 
^R \psi^{A,B\dagger}_\pm(x) = \psi^{A,B\dagger}_\pm(Rx), \nonumber \\
T: & \quad ^T \psi^{A,B}_\pm(x) = - \psi^{A,B\dagger}_\pm(Tx), \quad 
^T \psi^{A,B\dagger}_\pm(x) = \psi^{A,B}_\pm(Tx).
\end{align}

\subsection{Fermion Fields in Momentum Space Pockets}

As we know from the microscopic model, the electrons live in a momentum space 
pocket around $p = 0$ and have a spin opposite to the total magnetization,
while the holes live in a pocket around $p = \pi/a$ and have a spin parallel
to the magnetization. In order to describe these low-energy fermion degrees of
freedom, we perform a discrete Fourier transform from the sublattice 
indices $A$ and $B$ to the momentum space pocket indices $0$ and $\pi$. Again
this is in complete analogy to the antiferromagnetic case discussed in
\cite{Bru06,Bru07},
\begin{eqnarray} 
&&\psi^0_-(x) = \frac{1}{\sqrt{2}}\left[\psi^A_-(x) + \psi^B_-(x) \right], \quad
\psi^{0\dagger}_-(x) = \frac{1}{\sqrt{2}} \left[\psi^{A\dagger}_-(x) + 
\psi^{B\dagger}_-(x) \right], \nonumber \\
&&\psi^\pi_+(x) = \frac{1}{\sqrt{2}}\left[\psi^A_+(x) - \psi^B_+(x) \right], 
\quad
\psi^{\pi\dagger}_+(x) = \frac{1}{\sqrt{2}} \left[\psi^{A\dagger}_+(x) - 
\psi^{B\dagger}_+(x) \right].
\end{eqnarray}
We obtain the transformation rules
\begin{align}
SU(2)_s: & \quad \psi^0_-(x)' = \exp(- i \alpha(x)) \psi^0_-(x), \quad 
\psi^{0\dagger}_-(x)' = \exp(i \alpha(x)) \psi^{0\dagger}_-(x), \nonumber \\
& \quad \psi^\pi_+(x)' = \exp(i \alpha(x)) \psi^\pi_+(x), \quad 
\psi^{\pi\dagger}_+(x)' = \exp(- i \alpha(x)) \psi^{\pi\dagger}_+(x),
\nonumber \\
U(1)_Q: & \quad ^Q\psi^0_-(x) = \exp(i \omega) \psi^0_-(x), \quad 
^Q\psi^{0\dagger}_-(x) = \exp(- i \omega) \psi^{0\dagger}_-(x), \nonumber \\
& \quad ^Q\psi^\pi_+(x) = \exp(i \omega) \psi^\pi_+(x), \quad 
^Q\psi^{\pi\dagger}_+(x) = \exp(- i \omega) \psi^{\pi\dagger}_+(x), \nonumber \\
D: & \quad ^D\psi^0_-(x) = \psi^0_-(x), \quad 
^D\psi^{0\dagger}_-(x) = \psi^{0\dagger}_-(x), \nonumber \\
& \quad ^D\psi^\pi_+(x) = - \psi^\pi_+(x), \quad 
^D\psi^{\pi\dagger}_+(x) = - \psi^{\pi\dagger}_+(x), \nonumber \\
R: & \quad ^R\psi^0_-(x) = \psi^0_-(Rx), \quad 
^R\psi^{0\dagger}_-(x) = \psi^{0\dagger}_-(Rx), \nonumber \\
& \quad ^R\psi^\pi_+(x) = \psi^\pi_+(Rx), \quad 
^R\psi^{\pi\dagger}_+(x) = \psi^{\pi\dagger}_+(Rx), \nonumber \\
T: & \quad ^T\psi^0_-(x) = - \psi^{0\dagger}_-(Tx), \quad 
^T\psi^{0\dagger}_-(x) = \psi^0_-(Tx), \nonumber \\
& \quad ^T\psi^\pi_+(x) = - \psi^{\pi\dagger}_+(Tx), \quad 
^T\psi^{\pi\dagger}_+(x) = \psi^\pi_+(Tx).
\label{trans2}
\end{align}
The effective Lagrangian to be constructed in the next subsection must be 
invariant under all these symmetry transformations as well as under the
$SU(2)_Q$ transformations. The latter do not have a simple form in terms of the
momentum space pocket fields (and have thus not been listed here), but they
follow from eq.(\ref{transfderivations}).

\subsection{Effective Action for Charge Carriers}

We now construct the leading terms of the effective action for charge carriers,
which must be invariant under the symmetries $SU(2)_s$, $SU(2)_Q$, 
$D$, $R$, and $T$. We use the indices $n_t$, $n_x$, and $n_\psi$ in a
contribution to the Lagrangian
$\Lag_{n_t,n_x,n_\psi}$ to denote the number of temporal derivatives $n_t$, the 
number of spatial derivatives $n_x$, and the number of fermion fields $n_\psi$.
The  effective Lagrangian then takes the form
\begin{equation}
\Lag = \sum_{n_t,n_x,n_\psi} \Lag_{n_t,n_x,n_\psi}.
\end{equation}
The mass term is given by 
\begin{equation}
\Lag_{0,0,2} = M (\psi^{\pi\dagger}_+ \psi^\pi_+ - \psi^{0\dagger}_- \psi^{0}_-).
\end{equation}
In order to express the terms with spatial or temporal derivatives we introduce
the covariant derivative $D_\mu$ which acts as
\begin{eqnarray}
D_\mu\psi^0_-(x) = [\partial_\mu - i v_\mu^3(x)] \, \psi^0_-(x), \quad 
D_\mu\psi^{0\dagger}_-(x) = [\partial_\mu + i v_\mu^3(x)] \,
\psi^{0\dagger}_-(x), 
\nonumber \\
D_\mu \psi^\pi_+(x) = [\partial_\mu + i v_\mu^3(x)] \, \psi^\pi_+(x), \quad 
D_\mu \psi^{\pi\dagger}_+(x) = [\partial_\mu - i v_\mu^3(x)] \,
\psi^{\pi\dagger}_+(x).
\end{eqnarray}
Using the transformation laws of $v_\mu^3(x)$ listed in eq.(\ref{vmutrans}), one
arrives at the terms
\begin{equation}
\Lag_{1,0,2} = \psi^{\pi\dagger}_+ D_t \psi^\pi_+ + \psi^{0\dagger}_- D_t \psi^0_-,
\end{equation}
as well as
\begin{equation}
\Lag_{0,2,2} = \frac{1}{2M'} (D_1 \psi^{\pi\dagger}_+ D_1 \psi^\pi_+ - 
D_1 \psi^{0\dagger}_- D_1 \psi^0_-) + 
N (\psi^{0\dagger}_- v_1^- v_1^+ \psi^0_- +
\psi^{\pi\dagger}_+ v_1^+  v_1^- \psi^\pi_+).
\end{equation}
In contrast to an antiferromagnet, there is no fermion-single-magnon vertex.
Instead, all vertices contain at least two magnons. This implies that the
fermion-magnon interactions in a doped ferromagnet are of higher order than in
an antiferromagnet. 

Using the algebraic manipulation program FORM, we have also constructed all 
terms involving four fermion fields and up to one temporal or two spatial 
derivatives. They are not very illuminating and we thus do not list them here.
Instead we just concentrate on the fermionic Lagrangian in the two-hole sector,
which will be used later and which takes the form
\begin{equation} 
\label{leadingaction}
{\cal L} =  M \psi^{\pi\dagger}_+ \psi^\pi_+ + \psi^{\pi\dagger}_+ D_t \psi^\pi_+ +
\frac{1}{2M'} D_1 \psi^{\pi\dagger}_+ D_1 \psi^\pi_+ +
N \psi^{\pi\dagger}_+ v_1^+  v_1^- \psi^\pi_+ + 
G \psi^{\pi\dagger}_+ \psi^\pi_+ D_1 \psi^{\pi\dagger}_+ D_1 \psi^\pi_+,
\end{equation}
where $G$ is a 4-fermion coupling constant. It should be noted that the
effective coupling constants $M, M', N$ and $G$ are real. Since in the
above Lagrangian we have omitted the electron degrees of freedom, it is no
longer $SU(2)_Q$-invariant. Interestingly, the low-energy effective
Lagrangian has an emergent Galilean boost symmetry, despite the fact that the
underlying microscopic model does not possess this invariance.

\subsection{Determination of the Fermion Mass Parameters}

We now like to match the fermion mass parameters to the parameters of the
underlying microscopic system. Due to the $SU(2)_Q$ symmetry the masses of
electrons and holes are identical. Here we concentrate on the holes whose
dispersion relation is given by
\begin{equation} 
E_h(p) = M + \frac{p^2}{2M'},
\end{equation}
with the rest mass $M$ and the kinetic mass $M'$. In the microscopic model in 
eq.(\ref{energyH}) we calculated the dispersion relations of holes
\begin{equation} 
\label{lH}
E_h(p) = \frac{J}{2} + \frac{U}{2} - 2 t + t a^2 (p - \pi/a)^2 + 
{\cal O}(((p - \pi/a)^4).
\end{equation}
One should keep in mind that the holes live in momentum space pockets centered
at $p = \pi/a$, which must be taken into account in the matching of the
parameters. Indeed, we had already identified the rest and kinetic masses as
\begin{equation}
M = \frac{J}{2} + \frac{U}{2} - 2 t, \quad
M' = \frac{1}{2 t a^2}.
\end{equation}

\section{Non-perturbative Solution of the Effective Theory in the
Two-Hole Sector}

In the previous section we have constructed the effective field theory for
magnons and charge carriers, and we have fixed some of its low-energy
parameters by matching to the underlying microscopic model at half-filling as
well as in the single-hole sector. While the calculations in the microscopic
model were non-perturbative, until now the corresponding calculations in the
effective theory were based on perturbation theory. Indeed, it is a big
advantage of the effective theory approach to Goldstone boson physics that
perturbation theory provides quantitatively correct results in a systematic
low-energy expansion. While Goldstone bosons are derivatively and thus weakly
coupled at low energies, the contact interactions between two holes may very
well be strong, thus requiring a non-perturbative treatment not only of the
microscopic model, but also of the effective field theory. 

A similar situation arises in the effective field theory approach to the
strong interactions between nucleons and pions --- the Goldstone bosons of the
spontaneously broken chiral symmetry of QCD. There, the short-range
repulsion between two nucleons is again strong, which implies that the
effective theory must be treated non-perturbatively. In that case, it is still
an unsettled theoretical question how this can be achieved fully
systematically. In particular, there are various power-counting schemes, due
to Weinberg \cite{Wei90}, as well as due to Kaplan, Savage, and Wise
\cite{Kap98}, which are both not fully satisfactory. Recently, an
interesting modification of the Kaplan-Savage-Wise scheme has been proposed
\cite{Bea08}, and it remains to be seen whether this will finally resolve this
issue. In contrast to QCD or doped antiferromagnets, the ferromagnetic model
studied here has the advantage that it can be solved analytically. Hence,
one may reach a deeper understanding of the subtle non-perturbative fermion
dynamics. For this purpose, in this section we will investigate the two-hole
sector in the effective field theory and will then again compare with the
analytic results of the underlying microscopic model. 

\subsection{Solution of  the Two-Hole Schr\"odinger Equation}

In order to calculate the bound- and scattering-states of two holes in the 
effective theory, one can derive a two-hole potential from the effective
Lagrangian
of eq.(\ref{leadingaction}). The 4-fermion contact term of strength $G$ gives
rise to a potential that is proportional to the second derivative of a 
$\delta$-function. Such potentials are ultraviolet divergent and require 
renormalization even in quantum mechanics. In order to avoid the corresponding
subtleties, it is more efficient to apply the technique of self-adjoint
extensions. In particular, it is then not even necessary to explicitly
construct the potential.

Since the effective theory has an emergent Galilean boost symmetry, we may
consider the two-hole system in its rest frame. Introducing the relative
coordinate $x$ between the two holes, the Schr\"odinger equation reduces to a
single particle equation with the reduced mass $M'/2$. For kinematical reasons,
the two holes cannot exchange magnons. Instead, they just experience their 
4-fermion contact interaction. Away from the contact point $x = 0$, the 
two-hole Schr\"odinger equation thus describes free particles and is simply
given by
\begin{equation}
- \frac{1}{M'} \partial_x^2 \psi(x) = E \psi(x).
\end{equation}
In the theory of self-adjoint extensions, 
contact interactions in 1-dimensional quantum mechanics are treated by
removing the contact point $x = 0$ from the physical space. The effect of the
4-fermion interaction is then represented by a boundary condition on the
wave function. The most general self-adjoint extension has four independent
parameters and is characterized by the boundary condition
\begin{equation} 
\left(\begin{array}{*{2}{c}} \psi(\epsilon) \\ \partial_x \psi (\epsilon) \\
\end{array} \right) = \exp(i \theta)
\begin{pmatrix} a & b \\ c & d \end{pmatrix}
\left(\begin{array}{*{2}{c}} \psi(- \epsilon) \\ \partial_x \psi (- \epsilon) \\
\end{array} \right).
\end{equation}
Here $a$, $b$, $c$, and $d$ are real numbers with the constraint $ad - bc = 1$,
and $\epsilon$ is an infinitesimal displacement from the point $x=0$. Since
our system is parity-invariant (against the reflection $R$), the situation 
simplifies further and one obtains
\begin{equation} 
\label{selfcond}
\left(\begin{array}{*{2}{c}} \psi(\epsilon) \\ \partial_x \psi (\epsilon) \\
\end{array} \right) =
\begin{pmatrix} a & b \\ c & a \end{pmatrix}
\left(\begin{array}{*{2}{c}} \psi(- \epsilon) \\ \partial_x \psi (- \epsilon) \\
\end{array} \right),
\end{equation}
i.e.\ $\theta = 0$ and $d = a$. Since we are dealing with fermions, the Pauli 
principle implies a parity-odd wave function obeying $\psi(- x) = - \psi(x)$. 
The boundary condition on the wave function then implies
\begin{equation}
\label{condition_bs}
\kappa \psi(\epsilon) + \partial_x \psi(\epsilon) = 0, \quad 
\kappa = \frac{a+1}{b}.
\end{equation}
It should be pointed out that the wave function will in general not be
continuous at $x=0$. Alternatively to the 4-fermion coupling $G$, the strength
of the two-hole contact interaction can be characterized by the parameter
$\kappa$. Relating $\kappa$ to $G$ would require the ultra-violet
regularization of the second derivative of a $\delta$-function potential. We
avoid this unnecessary step by matching the value of $\kappa$ directly to the 
parameters of the underlying microscopic model.

Let us first search for two-hole bound states (with $E < 0$). The wave
function then takes the form
\begin{equation}
\psi(x) = A \exp(- \kappa x), \quad x > 0, \quad \psi(-x) = - \psi(x), 
\end{equation}
which is indeed discontinuous at $x=0$. The corresponding wave function in the
microscopic model was calculated in eq.(\ref{wfcoord}) and (in the rest frame,
i.e. for $p=0$) is given by
\begin{equation}
g(x) = C_0 (-1)^{x/a} \left(\frac{8 t}{J} \right)^{x/a}.
\end{equation}
The oscillating factor $(-1)^{x/a}$ is not present in the effective field theory
because of the momentum shift of the effective hole fields which are located
near $p = \pi/a$ in the Brillouin zone. Matching the exponential decays, we
identify
\begin{equation}
\label{kappa}
\kappa = - \frac{1}{a} \log\left(\frac{8 t}{J} \right).
\end{equation}
In the effective theory, the bound-state energy is given by
\begin{equation}
E_B = - E = \frac{\kappa^2}{M'}.
\end{equation}
The corresponding expression in the microscopic model was calculated in 
eq.(\ref{Eb_mmf}) and is given by
\begin{equation}
\label{ebind}
E_B = \frac{J}{4} \left(1 - \frac{8t}{J}\right)^2 =  
\frac{J}{8 t a^2 M_h'} \left(1 - \frac{8t}{J}\right)^2.
\end{equation}
Here we have used the value $M_h' = 1/2 t a^2$ for the kinetic hole mass.
Hence, from this expression one would conclude that
\begin{equation}
\kappa = \frac{1}{a} \sqrt{\frac{J}{8t}} \left(1 - \frac{8t}{J}\right).
\end{equation}
This is consistent with eq.(\ref{kappa}) only when $J \approx 8t$, i.e.\ when
the binding energy of eq.(\ref{ebind}) is small. In fact, the two expressions
even coincide up to second order in the perturbation $\delta=1-8t/J$. The
effective field theory thus provides a correct description of the bound state
only when the binding is weak. This is not surprising. If two holes form a
bound state with a large binding energy, this bound state must be introduced
in the effective theory as an independent degree of freedom. Only when the
bound state resembles a weakly coupled ``molecule'' in which the constituent
holes can be identified as relevant low-energy degrees of freedom, the
effective field theory (without explicit bound state fields) is appropriate.
In this context, it is interesting to note that the kinetic mass of two holes
calculated in eq.(\ref{mkinhh}) was given by
\begin{equation}
M_{hh}' = \frac{J}{8 t^2 a^2}.
\end{equation}
Only for $J = 8t$ this corresponds to the sum of the kinetic masses of
two holes $2 M_h' = 1/ta^2$. This is consistent, because the emergent Galilean
boost invariance of the effective theory indeed implies this relation.

Finally, let us consider the scattering states of two holes (with $E > 0$). We
make the ansatz
\begin{equation} \label{scattering_EFT}
\psi(x) = A \exp(i k x) + B \exp(- i k x),
\end{equation}
insert it into eq.(\ref{condition_bs}), and find
\begin{equation} 
\label{relEFT}
\frac{A}{B} = \frac{k + i \kappa}{k - i \kappa}.
\end{equation}
Let us now compare this result with the one obtained in the microscopic model
given in eq.(\ref{relMMF}). For low energies, i.e.\ for $q \rightarrow \pi/a$,
one obtains
\begin{equation}
\frac{\widetilde A}{\widetilde B} = 
\frac{k + i (1 - 8t/J)}{k - i(1 - 8t/J)} + {\cal O}(k^2).
\end{equation}
Indeed, using the value of $\kappa$ given in (\ref{kappa}), we finally get
\begin{equation}
\frac{A}{B} = \frac{\widetilde A}{\widetilde B}.
\end{equation}
We conclude that also here the effective field theory makes correct 
predictions, provided that the energies of both the bound state and the
scattering states are small.

\section{Conclusions}

We have investigated a Hubbard-type model for a doped  ferromagnet. While 
this model does not provide a realistic description of actual ferromagnetic 
systems, since it can be solved completely analytically, it provides a
stringent test of the corresponding low-energy effective field theory for
magnons and doped electrons or holes. Similar effective theories have been
constructed for magnons and charge carriers in the antiferromagnetic precursors
of high-temperature superconductors. Since, in that case, the underlying
microscopic models cannot be solved analytically, the correctness of the
effective field theory can only be tested in Monte Carlo simulations. Indeed,
such tests provide excellent numerical evidence for the validity of the
effective field theory approach. In the ferromagnetic case discussed here, the 
exact agreement between the analytic results of the microscopic and the 
effective theory lends further support to the validity of the systematic 
low-energy effective field theory technique. In particular, we like to stress
once more that the basic principles behind the construction of the effective
theory are the same for ferro- and for antiferromagnets.

While in this work we have investigated bound and scattering states of
two holes, another case of interest concerns the interaction between a spin
wave and a hole. Indeed, in the microscopic theory one is then lead to a
Faddeev-type equation and it would be instructive to confront the microscopic
result with the effective theory prediction also for this case. 

Effective field theories are also being used in the description of light
nuclei. In that case, a low-energy effective field theory of pions and
nucleons must be solved non-perturbatively, and it is currently not completely
clear how to do this in a fully systematic manner, i.e. based on a consistent
power-counting scheme. In this context, it is interesting that the two-hole
sector of the ferromagnetic model discussed here can be solved
non-perturbatively both in the microscopic and in the effective field theory
treatment. Both approaches agree as long as the two-hole binding energy is
small. On the other hand, when the binding becomes strong, the bound state
should be described by an independent effective field. The analytically
solvable test case of the ferromagnet may also provide valuable insights into
the subtle power-counting issues that arise in the context of the strong
interactions.

\section*{Acknowledgements}

The authors would like to thank C.\ Br\"ugger and M.\ Pepe for contributions
at an early stage of this work. C.\ P.\ H. would like to thank the members of
the Institute for Theoretical Physics at Bern University for their hospitality
during a visit at which this project was completed. F.\ K.\ is supported by an
SNF young researcher fellowship. The work of C.\ P.\ H.\ is supported by
CONACYT grant No.\ 50744-F. The work of U. Gerber is supported in part by
funds provided by the Schweizerischer Nationalfonds. The ``Albert Einstein
Center for Fundamental Physics'' at Bern University is supported by the
``Innovations- und Kooperationsprojekt C-13'' of the Schweizerischer
Nationalfonds.

\begin{appendix}

\section{Integrals for the Gap Equation}

In order to solve the gap equations (\ref{ABC1}), we needed the integrals
$I_1$,$I_2$,...,$I_6$ of eq.(\ref{I1}). The aim of this appendix is to show
how these integrals can be evaluated. Since 
\begin{equation}
\alpha I_2 + \beta I_3 + \gamma I_1 = 1, \quad I_4 + I_5 = I_1,
\end{equation}
one only needs to do four integrals, for example $I_1, I_2, I_4$, and $I_6$. 
These four integrals can be evaluated by using the residue theorem. We now 
explicitly present the calculation for
\begin{equation} 
I_1 = \frac{1}{2 \pi} \int_{-\pi/a}^{\pi/a} dq \ 
[\alpha \cos(q a) + \beta \sin(q a) + \gamma]^{-1}.
\end{equation}
We integrate around the unit circle $C$, and thus make the substitution
\begin{equation}
z = \exp(iqa). \nonumber
\end{equation}
Then we can write
\begin{equation}
I_1 = \frac{1}{2 \pi i} \oint_{C} dz \
[\frac{\alpha}{2} (z^2 + 1) + \frac{\beta}{2 i} (z^2 - 1) + \gamma z]^{-1}.
\end{equation}
The integrand has two singularities at
\begin{equation}
z_A = \frac{- \gamma + \sqrt{\gamma^2 - \alpha^2 - \beta^2}}{\alpha - \beta i},
\quad
z_B = \frac{- \gamma - \sqrt{\gamma^2 - \alpha^2 - \beta^2}}{\alpha - \beta i}.
\end{equation}
We get a contribution to the integral only if the singularities lie within the
unit circle $C$. Therefore we consider the absolute values of $z_A$ and $z_B$,
\begin{equation}
|z_A|^2 = \frac{(\sqrt{\gamma^2 - \alpha^2 - \beta^2} - \gamma)^2}
{\alpha^2 + \beta^2}, \quad
|z_B|^2 = \frac{(\sqrt{\gamma^2 - \alpha^2 - \beta^2} + \gamma)^2}
{\alpha^2 + \beta^2}.
\end{equation}
We find
\begin{equation}
|z_A|^2 < 1 \Leftrightarrow \gamma > 0, \quad
|z_B|^2 > 1 \Leftrightarrow \gamma > 0, 
\end{equation}
as well as
\begin{equation}
|z_A|^2 > 1 \Leftrightarrow \gamma < 0, \quad
|z_B|^2 < 1 \Leftrightarrow \gamma < 0.
\end{equation}
Hence for $\gamma > 0$ only $z_A$ lies within the unit circle $C$ and for 
$\gamma < 0$ only $z_B$ lies within $C$. The residues of the poles at $z_A$ and
$z_B$ are given by
\begin{eqnarray}
&&R_A = \frac{- 2 \gamma + 2 \sqrt{\gamma^2 - \alpha^2 - \beta^2}}
{2 \alpha \sqrt{\gamma^2 - \alpha^2 - \beta^2} - 
2 i \beta \sqrt{\gamma^2 - \alpha^2 - \beta^2}}, \nonumber \\
&&R_B = \frac{2 \gamma + 2 \sqrt{\gamma^2 - \alpha^2 - \beta^2}}
{2 \alpha \sqrt{\gamma^2 - \alpha^2 - \beta^2} - 
2 i \beta \sqrt{\gamma^2 - \alpha^2 - \beta^2}}.
\end{eqnarray}
Collecting the results we obtain
\begin{equation}
I_1 = \mbox{sign}(\gamma) \frac{1}{\sqrt{\gamma^2 - \alpha^2 - \beta^2}}.
\end{equation}
The integrals $I_2$, $I_4$, and $I_6$ can be obtained completely analogously.

\end{appendix}

\end{document}